\documentclass[afterpage,useAMS,usenatbib]{mn2e}

\usepackage{psfig}

\def\mgs{M_{gas}/M_{stars}}
\def\rhalf{R_{\rm half}}
\def\rcore{R_{\rm core}}
\def\eg{{\it eg.\thinspace}}

\def\kms{{\rm\,km\,s^{-1}}}

\def\Msun{{\rm\,M_\odot}}

\def\pc{{\rm\,pc}}

\def\Myr{{\rm\,Myr}}

\def\srm1{{\rm\,sr^{-1}}}

\title[Core Dissolution and massive stars] {Core dissolution and the dynamics of massive stars in young stellar clusters}

\author[Vine \& Bonnell]{S.\ G.\ Vine \&\ I.\ A.\ Bonnell\\
School of Physics and Astronomy, University of St Andrews, North Haugh, St Andrews, Fife KY16 9SS}

\begin{document}

\date{\today}
\pagerange{\pageref{firstpage}--\pageref{lastpage}} \pubyear{2002}

\maketitle

\label{firstpage}

\begin{abstract}
We investigate the dynamical effects of rapid gas expulsion from the
core of a young stellar cluster. The aims of this study are to
determine 1) whether a mass-segregated core survives the gas expulsion
and 2) the probable location of any massive stars that have escaped
from the core.  Feedback from massive stars is expected to remove the
gas from the core of the cluster first, as that is where most massive
stars are located.  We find that gas expulsion has little effect on
the core for a core star formation efficiency, $\epsilon$, of greater
than 50\%. For lower values of $\epsilon$ down to 20\%, a reduced core
survives containing the majority of the massive stars while some of
them are dispersed into the rest of the cluster. In fact we find that
ejected stars migrate from radial to tangential orbits due to stellar
encounters once they leave the core. Thus, the location of massive
stars outside of the core does not exclude their forming in the dense
cluster core.  Few massive stars are expected to remain in the core
for $\epsilon$ lower than 20\%.

\end{abstract}

\begin{keywords}
stars: formation -- stars: massive  -- stars: luminosity function, mass function -- globular clusters
and associations: general.
\end{keywords}

\section{Introduction}

The origin of massive stars is a poorly understood phenomena.  One
clue to their formation is that they are generally found in the dense
cores of young stellar clusters (\citealt{hill:1997}; \citealt{cs:1998};
\citealt{cbh:2000}; \citealt{mg:2001}; \citealt{zmw:1993}). Their location in 
these very young systems cannot generally be explained by dynamical
mass segregation (\citealt{bd:1998}). Higher than observed primordial
core densities (\citealt{kroupa:2002}; \citealt{kah:2001}) can cause
rapid mass segregation in the core region.
 The central location
of massive therefore must tell us something about their birth.  One
possibility is that the accretion rates in the centre of the cluster
are higher due to the deeper potential wells (\citealt{bbcp:2001};
\citealt{bonnell:2002}) and that therefore the protostars that are
present in the core are able to accrete up to a high mass (\eg
\citealt{bm:2001}, \citealt{klessen:2001}). An additional possibility
is that as the stars in the core accrete, the core contracts due to
the increase in the potential energy. The increase in stellar density
may then be sufficient such that collisions between intermediate mass
stars occur to form the massive stars (Bonnell, Bate, \& Zinnecker
1998; \citealt{bonnell:2002}).  One of the successes of these models
is that they naturally account for the formation of massive stars in
the centres of stellar clusters. These models predict the location of
massive stars at the end of the accretion, mass-buildup phase.  At
this point the cluster contains significant mass in gas which can
still affect the cluster dynamics.

Once a massive star has formed in a cluster, it can have dramatic
effects on its surroundings. The strong winds and ionising photons
inject significant energy into the surrounding gas which will begin to
expand, eventually being cleared from the cluster. This gas removal
has two effects: firstly, it halts any residual accretion or star
formation; secondly, if the gas still comprises a significant fraction
of the total mass, the cluster will expand to reflect the new, lower
potential. Simulations of this process (\citealt{gb:2001};
\citealt{kah:2001}; \citealt{kb:2002}) have concentrated on the global
evolution of the cluster as the gas is removed.  They find that if the
gas is a small fraction of the total mass, or if it is removed slowly
relative to the dynamical timescale, then the cluster can retain
significant fraction of its initial stars. \cite{kah:2001} suggests
that early dynamical mass segregation of the most massive stars can
survive during and after a gas expulsion phase.  \cite{kroupa:2000}
has presented results specifically for models of the Orion Nebula
Cluster (ONC) including the gas expulsion scenario reflecting the
situation we model here, but without primordial mass segregation.

In this paper, we are concerned with the dynamical evolution of the
cluster core when the gas is removed. Gas removal occurs from inside
out as the massive stars inject energy into their environment. The
core of the cluster will be cleared first and the core stars will be
affected by the gas removal before the rest of the cluster. This has
serious implications for the dynamics of the massive stars, their
location in the cluster, or even their possible ejection from the
cluster. We address these questions using numerical simulations
presented below.

\section{Gas expulsion from core}
Gas removal occurs due to the injection of energy from the massive
stars in the core. The energy produced by these O and B type stars can
take the form of strong stellar winds of up to a few $\times 1000$ km
s$^{-1}$ (\citealt{churchwell:1999}) or the formation of an expanding
HII region ($v_{HII} \approx 10$ km s$^{-1}$). In both cases, the
expansion velocity is significantly higher than the velocity
dispersion seen in young stellar clusters due to the higher initial
density.  If the core of the cluster is about 0.1pc this means that
the crossing time {\em for the core} $(t_{\rm
cross}\sim5\times10^4yr)$ is at least an order of magnitude larger
than the gas expulsion time.  We may then assume that the period of
gas expulsion has negligible effect on the dynamics of the constituent
stars in the core, and so we assume instantaneous gas expulsion.  Thus
we incorporate an initial surplus of kinetic energy into the core
stars which results in their being no longer in virial
equilibrium. This implies that the gas is only removed from the core
as stars further out remain in virial equilibrium. The timescales
$(\sim10^5yr)$ and radial extent $(\sim0.1pc)$ reflect the situation
in an ultra-compact HII region where a stellar wind has evacuated a
small inner region before stalling due to mass loading
(\citealt{lizano:1996}), ram pressure of the infalling gas
(\citealt{wc:1989}), or other possible mechanisms
(\citealt{churchwell:1999}).

\section{Numerical Simulations}

Over the timescales involved it is only gravitational forces which
play a significant r\^ole in the subsequent evolution.  Thus the
system is sufficiently modelled with a pure N-body simulation.  The
numerical simulations reported here were performed using the {\em
NBODY2} code (\citealt{aarseth:2001}). This code is extremely
efficient and accurate at following a stellar dynamical system for
tens to a hundred dynamical times. It includes the AC neighbour scheme
(\citealt{ac:1973}) and a softened potential (here $r_{\rm
soft}\la0.1\times r_{\rm core}$) which minimises computational expense
at the cost of not following any (close) binary systems. As we are
following the stellar dynamics over only a few local dynamical times
where distant interactions dominate, {\em NBODY2} is sufficient for
our needs.
 
\begin{figure}
  \psfig{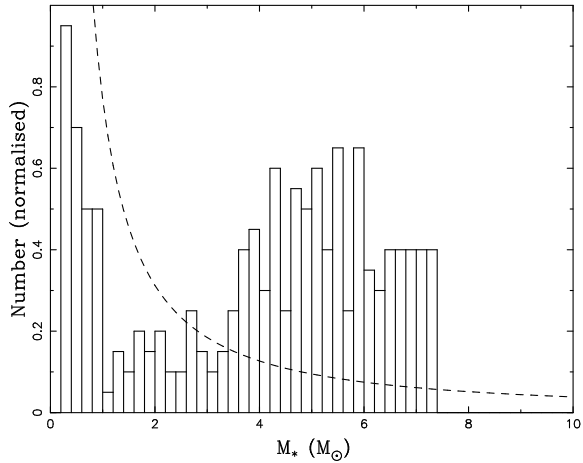} \caption{The mass
  distribution of the stars in the core. This is an average of all initial
  realizations. This demonstrates the initial mass segregation,
  compared to the non-mass segregated initial $\alpha = 2.3$
  distribution shown here as a dotted line.}  \label{initcore}
\end{figure}

Our cluster is assumed to be in a virialised state prior to gas
expulsion, from arguments of the star formation timescale compared to
the crossing times, typical systems may indeed be near to viral
equilibrium (\citealt{kb:2002}). The stellar system consists of 1500
stars populated with masses distributed according to a Salpeter mass
function, \hbox{$\phi(m)\propto m^{-2.3}$}, ranging from $0.1-10\Msun$
and distributed in phase space with a plummer density
profile. Additionally, it is necessary to model an initially mass
segregated system such that the massive stars are found in the core
region of the cluster, as described above. We construct such initial
conditions by allowing our randomly distributed stellar system to
relax for several dynamical times until the massive stars have sunk to
the core.  This system is then used as our initial condition for the
simulations in this investigation.  It must be emphasised that this
process of achieving initial conditions is only to construct a model
of the observed initial state and not to represent the physical method
by which massive stars end up in the cores of very young stellar
clusters in the first place. Our pre-initial clusters are evolved
until little more mass segregation can occur.  The resulting initial
mass distribution in the core is shown in Figure~\ref{initcore}
showing that the high mass stars have sunk to the core.

\begin{figure*}
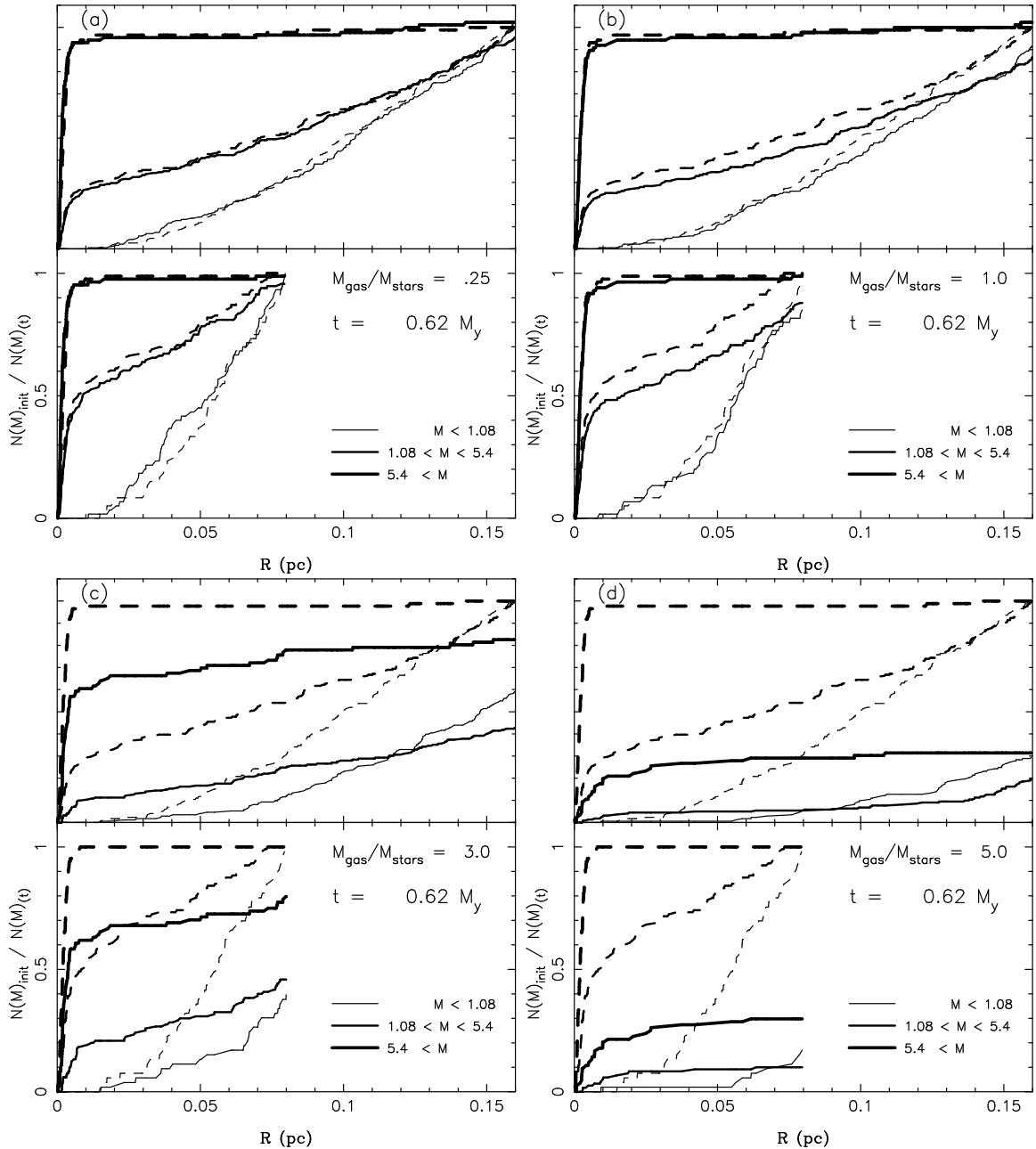

  \hbox{\psfig{figure=c25.ps,angle=270,width=3.0in}
  \psfig{figure=c1.ps,angle=270,width=3.0in}}
  \hbox{\psfig{figure=c3.ps,angle=270,width=3.0in}
  \psfig{figure=c5.ps,angle=270,width=3.0in}} \caption{Cumulative
  number ratio for different mass ranges of stars, and different
  initial gas ratios averaged over all realisations.  The vertical
  axis shows the cumulative numbers of stars divided by the total
  number contained within the region of interest.  Radius is shown
  along the horizontal axis.  Three mass ranges are shown, indicated
  by a thick line $(M_*\ge 5.4\Msun)$, a medium line
  $(5.4\Msun>M_*>1.08\Msun)$, and a thin line $(M_*\le1.08\Msun)$.
  The dashed lines represent the initial state of the clusters for
  each mass range.  The lower panel in each figure shows the
  cumulative number ratio normalised to the initial numbers of stars
  within $\rcore$, and the upper panel within $2\rcore$. The upper
  figures, (a) $\mgs=0.25$ and (b) $\mgs=1.0$, show little evolution;
  figure (c) $\mgs=3.0$ shows significant numbers of stars lost from
  the core; and figure (d) $\mgs=5.0$ show 70\% of the massive stars
  are lost from within $\rcore$.}  \label{cumdist}
\end{figure*}

\begin{table}
\centering
\begin{tabular}{ccc}
$\frac{M_{gas}}{M_{stars}}$ & $\epsilon$\\
\hline
0.25 	&	80\% \\
1.0  	&	50\% \\
2.0	&	33\% \\
3.0	&	25\% \\
4.0	&	20\% \\
5.0	&	17\% \\
\end{tabular}
\caption{Values of the initial gas ratio parameter used, and the equivalent star formation efficiency, $\epsilon$.}
\label{gf}
\end{table}

Our results are presented scaled to the physical units which apply to
the Orion nebula cluster as taken from \cite{hh:1998}, it should be
emphasised that our models are not intended to model the ONC which has
a much wider mass range $(0.1-50\Msun)$.  The half mass radius of the
cluster, $r_{\rm half}=0.8{\rm pc}$. We choose the core radius to be
$r_{\rm core}=0.08\pc$ being 0.1 the half mass radius, it is this
region within $r_{\rm core}$ in which we consider the gas expulsion to
be effective. The mass of our cluster is $1200\Msun$, and N$=1500$ so
that the mean stellar mass, $<{\rm\,M_*}>=0.8\Msun$.  Finally, we
follow the evolution of the core stars to $t\ga 1.0\Myr$
$(\sim20t_{\rm cross})$, by which time the dynamical evolution we are
interested in will have occurred.

\section{Run Parameters}
The effect of gas removal is included by way of a super-virial
distribution of the stellar velocities inside the cluster's core. The
kinetic energy of these stars is taken to be that required to balance
an additional mass of gas, $M_{gas}$. We have parametrised our model
by the ratio of $M_{gas}/M_{stars}$ prior to gas expulsion in the
core. The values of this parameter used and the equivalent star
formation efficiencies are shown in Table~\ref{gf}. This method mimics
the case when gas removal is instantaneous. This is the worst case
scenario for retaining the stars and thus gives a conservative
estimate of how many massive stars will be retained in the core.

In practice we take our initial system as described above and increase
the kinetic energies of the stars in the core by the following energy
argument,
\begin{equation}
[M_{gas}+M_{stars}]v_{old}^2 = M_{stars}v_{new}^2
\end{equation}
we can then directly scale the core velocities such that,
\begin{equation}
v_{new}^2 = \frac{v_{old}^2}{\epsilon}
\end{equation}
where we have introduced the star formation efficiency, in terms of the gas ratio as, 
\begin{equation}
\epsilon = 1/(1+\frac{M_{gas}}{M_{stars}}).
\end{equation}

\begin{figure}
  \psfig{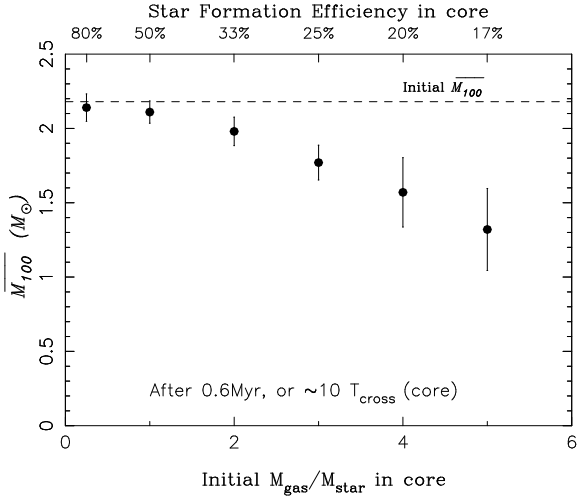} 
  
  \caption{The figure shows the average mass of the 100 stars nearest
  the centre of mass of the cluster for each value of $\mgs$.  The
  vertical bars show the resulting range within the 8 different
  realisations of each run.  A roughly linear relation between $\mgs$
  and $\overline{M_{100}}$ after $\mgs=1.0$.}

\label{mmass}
\end{figure}
\begin{figure*}
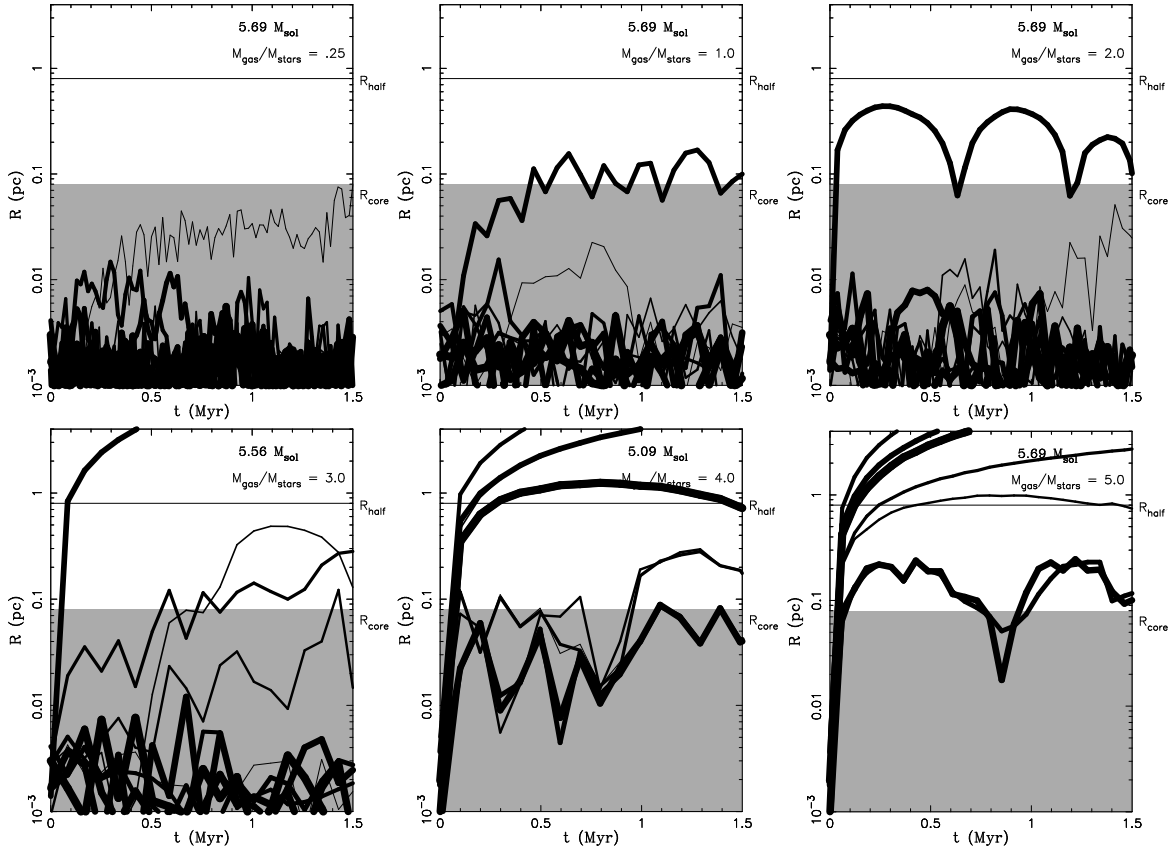

  \hbox{\psfig{figure=r25d.ps,width=2.0in,angle=270}
  \psfig{figure=r01g.ps,width=2.0in,angle=270}
  \psfig{figure=r02d.ps,width=2.0in,angle=270}}
  \hbox{\psfig{figure=r03d.ps,width=2.0in,angle=270}
  \psfig{figure=r04g.ps,width=2.0in,angle=270}
  \psfig{figure=r05g.ps,width=2.0in,angle=270}} 

  \caption{Radial distance from density centre vs time for the 10 most
  massive stars initially found in the cluster core. These figures
  represent a single individual realisation for each value of the
  parameter $\mgs$. The grey area is the core region in which the
  initial velocity kick is given, and a solid line representing the
  initial $\rhalf$ is shown.  The thickness of the individual lines
  relate to the mass of the star, thickest being most massive. The
  realisations have been chosen to illustrate the typical behaviour of
  the system for each $\mgs$, where no stars leave the core for a
  value of 0.25 and all leave the core for $\mgs=5.0$. The apparent
  kink in the path of stars that leave the core is due to the time
  sampling of the simulations and is not a physical affect. The value
  of the lower mass limit of stars shown is also indicated in each
  figure.}  
  \label{massrad}
\end{figure*}

  For each value of the initial parameter, $\mgs$, 8 initial
  realisations were constructed using a different random seed for
  each.  This helps to extract the mean behaviour of the massive stars
  in the clusters and not individual fluctuations due to their small
  numbers.

\section{Results}
The purpose of our investigation is to discover whether a mass
segregated core will survive after a gas expulsion phase, and also to
gain general insight to the subsequent motions of the massive stars.
The following results examine properties of the core: mass
segregation, average mass within the core, fraction of stars remaining
in the core, and core survivability.  We also look at the phase space
evolution of the individual massive stars by examining their radial
evolution and their angular momentum.

\subsection{Mass Segregation}
We investigate the evolution of mass segregation in the core, which,
in this case, amounts to the survival of the most massive stars in the
core.  

After $0.62\Myr$ $(\sim 12t_{\rm cross})$ we take the ratio of the
number of stars within $r$ to the number initially contained within
given radii ($\rcore$ and $2\rcore$).  This cumulative number ratio is
calculated independently for 3 stellar mass ranges: $M_*\le 1.08\Msun,\;
1.08\Msun<M_*<5.4\Msun, \; 5.4\Msun\le M_*$, shown as a thin line, a
medium line, and a thick line respectively in
Figures~\ref{cumdist}(a)--(d). These figures show the cumulative
number ratio for a $\mgs$ parameter of (a) 0.25, (b) 1.0, (c) 3.0, and
(d) 5.0.  The top and bottom panels in each figure show the cumulative
number ratio normalised to $2\rcore$ and $\rcore$ respectively.  The
initial state of the system at $t=0$ is shown as a dashed line in each
figure.  These figures are the combined results of all realisations
for each particular $\mgs$.

In particular, we are interested in the highest mass range, which we
consider to represent the OB type stars in the cluster.  Clusters with
lower values of $\mgs$ (figures~\ref{cumdist}(a) and (b)) show little
significant change in the distribution of the massive stars, in fact
the lower mass stars decrease in number more than the massive stars,
making the core slightly more segregated.  However, as soon as the
effective kick to the core stars becomes large enough to boot out the
massive stars we see a dramatic decrease in all the stellar numbers.
 Figures~\ref{cumdist}(c) and (d) show a decrease in
the actual massive star numbers by 20\% and 70\% respectively.

An alternative and perhaps more intuitive way of looking at the mass
segregation is to look at the average stellar mass within a certain
radius, or as we do in this case the 100 stars nearest to the centre
of mass. Figure~\ref{mmass} shows there is an almost linear decrease
in $\overline{M_{100}}$ for each value of the $\mgs$ parameter after
$\mgs=1.0$. Also shown, as a vertical bar on each point, is the
range due to the variation in the initial realisations for each
parameter.  This shows clearly that as the stars are ejected out into
the rest of the cluster, the massive stars are less of a
characteristic of the central cluster regions.  Although the mean
stellar mass is slightly greater than the cluster mean
$<M_*>=0.8\Msun$, at $\mgs=5.0$ the mass segregation is much less
significant.

\subsection{Radial evolution of Massive stars} 
The question now arises of what becomes of the massive stars which the
core has ejected.  For comparison with observed clusters we would like
an idea of where the stars are located in our simulations. We examine
the radial distances of the most massive stars from the centre of mass
with respect to time.  Typical examples are shown in
Figure~\ref{massrad}.
\begin{figure}
  \psfig{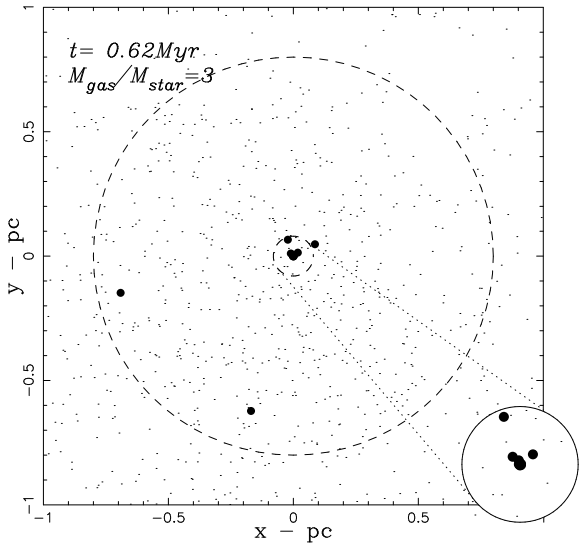} 
  \caption{The $x-y$ plot
  of a typical realisation with $\mgs=3$ after $t=0.62\Myr\;(\sim12t_{\rm cross})$.  Here we notice two stars
  remain bound within $\rhalf$ (large dashed circle) and the rest
  forming a bound (but less dense) core shown in the inset.}
  \label{xyplot}
\end{figure}

These figures illustrate typical evolution of the 10 most massive
stars which were initially in the cluster core, each figure showing
one realisation for each value of the parameter $\mgs$.  As we have
seen in the previous section, for a larger $\mgs$, a greater number of
massive stars leave the core.  Figure~\ref{massrad} shows this trend
explicitly and we can see the subsequent radial motion of the stars
after gas expulsion.  The grey region in the figures is the initial
core in which the gas is assumed to have been removed.  The half mass
radius, $\rhalf$, is also shown as a solid line at 0.8pc.

For a
$\mgs=0.25$, the figure shown top left is typical for all
realisations and similar results are seen for $\mgs=1.0$ (top middle). For
$\mgs=2.0$ (top right) it is more typical to see a couple
of the stars ejected from the core but within the half mass
radius. For $\mgs=3.0$, 2 to 4 stars are ejected beyond
$\rhalf$ and another couple beyond $\rcore$ (bottom left). 
Figure~\ref{xyplot} shows a resulting configuration in
the $x-y$ plane.  There is a resemblance to the configuration of the
massive stars in the Orion Nebula Cluster (ONC), showing that this is a possible mechanism
leading to ONC type structures, where some of the massive stars are located a significant distance from the core.  Although without observational
velocities we cannot say how likely it is.

The tendency when $\mgs=4.0$ (bottom middle) is for
more massive stars to be completely ejected and still two or so to be found
between $\rcore$ and $\rhalf$.  It is likely that most stars are
ejected from the core for $\mgs=5.0$ and also beyond $\rhalf$ 
(Figure~\ref{massrad}, bottom right).
\begin{figure}
	\psfig{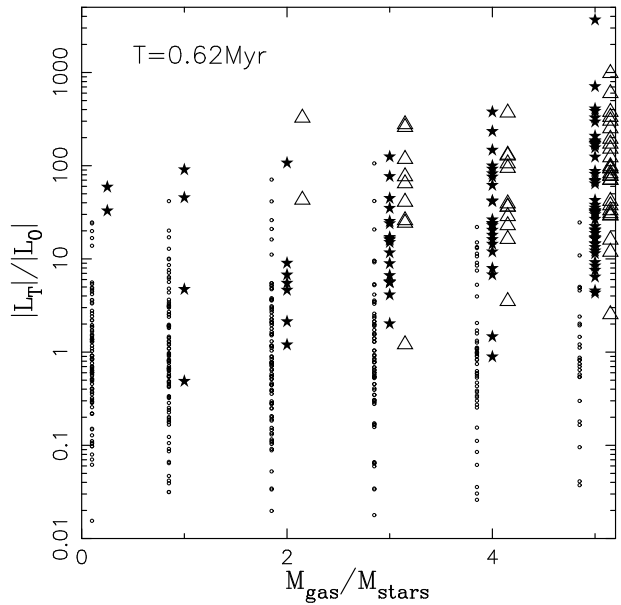} 

	\caption{ Ratio of angular momentum $|\bmath{L}_T|$ at
	$t=0.62\Myr (12t_{\rm cross})$ to initial angular momentum
	$|\bmath{L}_0|$ for all stars over $5\Msun$. All the data for
	each realisation of each $\mgs$ is shown, separated into stars
	which are still in the core at $t=0.62\Myr$ ($\circ$, shifted
	left for clarity), stars beyond the core $(\star)$, and stars
	which have left the cluster ($\triangle$ shifted right, for
	$r>5\rhalf$).  The distribution shows a distinct bimodality
	between core stars which show little overall increase in
	$|\bmath{L}|$, and stars that have left the core which all
	show an increase in $|\bmath{L}|$ becoming greater with higher
	$\mgs$}

	\label{lratio}
\end{figure}  
\subsection{Velocity Evolution}
Observational measurements of velocities (proper motions as well as
line-of-sight velocities) are now feasible with space borne
instruments, such velocities may provide strong evidence to decide
on the spatial origin of the stars. It is therefore of interest to explore the kinematics that result from gas expulsion and the ejection of massive stars from the core of the cluster. 

The expectation is that stars which receive significant velocity
increases escape the cluster on near-radial orbits. Instead, we show
below that these stars generally have more tangential than radial
motions. This can be explained by the interactions that occur due to
the discrete nature of the stellar cluster.  We can suppose that the
orbits of these stars will encounter a great range of stars on highly
varied orbits, its velocity signature will vary greatly, leaving the
angular momentum measured with respect to the centre of the core
density peak, $|\bmath{L}|$, as the most robust general measure of the
stars' subsequent evolution.  This measure of angular momentum is not
affected by the motion of core over the duration of these simulations.

Figure~\ref{lratio} shows the ratio of the angular momentum at time
$t=0.62\Myr (12t_{\rm cross})$ to the initial value immediately after
the velocity kick $|\bmath{L}_T|/|\bmath{L}_0|$. All stars over
$5\Msun$, in all realisations of the parameter $\mgs$ are shown.  We
have distinguished the stars by their radial position at time
$t=0.62\Myr$: those within $\rcore$ are shown as small open circles
($\circ$), those beyond $\rcore$ shown as filled stars ($\star$), and
those beyond $5\rhalf$ (those which have left the cluster) are shown
as open triangles.  Each radial range is shifted slightly to left or
right for clarity.

The first effect to notice is the increasing spread of angular
momentum with increasing initial $\mgs$, such that for $\mgs=5.0$ the
gain in $|\bmath{L}|$ for some stars reaches $\sim1000$. The other
striking feature is the bimodality between stars that have stayed in
the core and stars which have left it. Stars remaining in the core have angular
momenta distributed around their initial values, although plenty of
angular momentum exchange has occurred resulting in
$100>|\bmath{L}_T|/|\bmath{L}_0|>0.01$. The core escapers, however, almost
exclusively increase their angular momenta such that
$1000>|\bmath{L}_T|/|\bmath{L}_0|>1$.  

Such a result indicates that the higher velocity massive stars must
have encountered a greater number of stars with a broader range of
orbits on their journey outward through the cluster.  A combination of
their gain in $|\bmath{L}|$ and high velocity puts many stars on large
tangential orbits which do not pass through the core.  This was seen in
Figure~\ref{massrad} as stars are ejected from the core but are seen
to orbit beyond $\rcore$.  As a consequence of this the velocities of
the ejected massive stars will be seen to have predominantly
tangential motions to the cluster core.

There is a further subset of stars which appear to leave the cluster
entirely (a limit of $5\rhalf$ is used to identify cluster leavers),
these are shown as open triangles in Figure~\ref{lratio}.  The
physical velocities of these stars range from $10$ to $40\kms$.  A
result which allows us to rule out the existence of observed high
velocity stars ($\sim200\kms$) as a consequence of the two body
interactions which occur in this scenario.

\subsection{Core Survival}
\begin{figure}
  \psfig{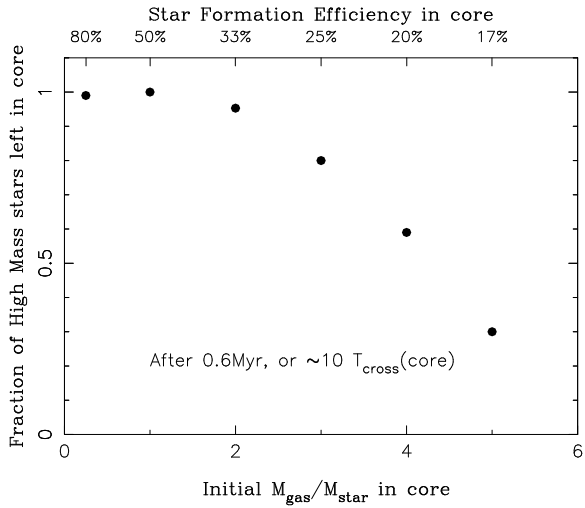}
  \caption{Quantifying core dissolution for different initial $\mgs$ values, and the equivalent star formation efficiencies, after a period of 0.62Myr $(12t_{\rm cross})$}
  \label{survival}
\end{figure}

To summarise the results in terms of the original question posed ---
how does gas expulsion from the core affect the core's survival? ---
we have plotted (Fig~\ref{survival}) the fraction of massive stars
remaining in the core for each value of $\mgs$.  It is perhaps a matter
of definition as to what constitutes a surviving core, and in this
case we take a value of 50\% of the original massive stars. More
precisely we say that if more than half the core (or rather its
constituent massive stars) is dissipated subsequent to the gas
expulsion phase then the core does not survive. This is supported by
the fact that the actual number of massive stars in the core are of
the order of a few, once half of these have left the system there may
be only one or two left, not enough to constitute a core.

The relation between $\mgs$ and the core survivability is fairly
smooth (figure~\ref{survival}). The core is relatively undisturbed up
to $\mgs=2.0$, then becomes proportionately more disrupted with
greater $\mgs$ but survives (by our definition) to a $\mgs$ of
over $4.0$.

\section{Discussion}
We have modelled the gas expulsion from the core of a stellar cluster
via a rapid change in the stellar velocities and thus an effective
change in the gravitational potential of the core. This is taken to
mimic the evolutionary stages of a hyper or ultra-compact HII region.
Their evolution is likely to be driven by a stellar wind which stalls
as mass loading occurs in order to explain their relatively long
lifetimes $(\sim10^5yr)$.  Thus, we assume an instantaneous removal of
gas from the core of the cluster but assume that the gas content of
the rest of the cluster is for the most part unaffected during our simulation. 

The formation of massive stars is largely an unsolved problem. What is
certain is that they require high accretion rates in order to build
the star up over timescales of $10^5$ to $10^6$ years. This is true
whether they form just through accretion (\citealt{bm:2001};
\citealt{bonnell:2002}) or even through stellar mergers which require
high accretion rates to reach the necessary stellar density
(\citealt{bbz:1998}; \citealt{bonnell:2002}). The most obvious way
that high accretion rates can occur is by situating the forming
massive star in the centre of a stellar cluster. In this way, the
overall cluster potential funnels matter down to the protomassive star
where it can preferentially accrete it due to its position and mass
(\citealt{bbcp:2001}). This naturally results in a mass segregated
stellar cluster even while deeply embedded.

The simulations reported here show that gas expulsion is unlikely to
remove this 'initial' mass segregation unless the gas comprises more
than 83 per cent of the total mass {\it in the core}. The core, being
where the accretion rates are highest and the most massive stars are
located, has therefore the lowest gas fraction of the system and is the most
likely to be dominated by the mass in stars. Thus, the mass
segregation which results from the cluster formation and gas accretion
is likely to remain unaffected by the gas expulsion stage in the
cluster.

In some young clusters such as the ONC, a few relatively massive stars
are found to be some distance from the core of the cluster. If the
above scenarios for the formation of massive stars are correct in that
they form in the denser parts of the cluster, then we need to explain
how they could have moved from the core to, for example, the half-mass
radius of the cluster. the numerical simulations reported here offer
one possibility.  The massive stars could have originated in the core
but escaped from this region during the gas expulsion phase (\citealt{w:2002}). We see
that for gas fractions of greater than 83 per cent, some of the
massive stars do escape the core and can be found virtually anywhere
in the cluster. They will of course sink back to the cluster core but
on longer timescales than the age of the cluster (\citealt{bd:1998}).
 
The difficulty in the above is that relatively high gas fractions are
required which may not be present in the core of the cluster. An
alternative mechanism to explain the location of the massive stars not
in the core would be stellar interactions with a binary or multiple
system which eject one or more of the stars. Some of these ejections
could be the high-velocity OB runaways while those that do not garner
such a large velocity boost remain in the cluster but outside the
core. If that is the case, their dynamics should resemble the massive
stars ejected from the core in the simulations reported here.
   
\section{Conclusion}
Using a series of N-body simulations we follow the dynamical evolution
of massive stars which are assumed to be initially located in the core of the
cluster. The stars undergo evolution because of very rapid gas
expulsion (assumed instantaneous) which is parametrised by the ratio
of gas mass to stellar mass in the core $(\mgs)$.  This parameter is
varied from 0.25 to 5.0 (equivalent to a star formation efficiency of
80\% down to 17\%).

We find that the initial mass segregation in the core is destroyed for
an initial gas mass of greater than 4 times the stellar mass, or a
star formation efficiency of less than 20\%. The initial conditions
used are of a relaxed, mass-segregated cluster. The massive are
therefore approximately in equipartition with low-mass stars in the
core.  Their lower velocities result in a lower probability of
escaping the core once the gas too has been removed.  This explains the
relatively high value of $\mgs$ required to dissolve the core.
Otherwise, for star formation efficiencies of more than 20\%, but less
than 50\% some massive stars will be ejected from the core and maybe
beyond. However these ejected stars are found to have predominantly
tangential motions, gaining angular momentum due to encounters along
the way.  


\bibliographystyle{mn2e} 
\bibliography{cdiss} 

\bsp
\label{lastpage} 
\end{document}